\documentclass{aastex631}
\usepackage[toc,page]{appendix}
\usepackage[caption = false, position=top]{subfig}
\usepackage{amsmath}
\usepackage{amssymb}
\usepackage[normalem]{ulem}
\usepackage{soul}
\usepackage{booktabs}
\usepackage{sidecap}
\usepackage{floatrow}
\usepackage{mathtools}
\usepackage{tgcursor}

\newcommand{\bxi}{{\boldsymbol\xi}}
\newcommand{\bx}{{\boldsymbol x}}
\newcommand{\bnabla}{{\boldsymbol \nabla}}
\newcommand{\bcurl}{{\boldsymbol \times}}
\newcommand{\bfu}{{\boldsymbol u}}

\newcommand{\pol}{P}
\newcommand{\tor}{T}
\newcommand{\bq}{{\boldsymbol q}}

\newcommand{\bk}{{\boldsymbol k}}
\newcommand{\bL}{{\boldsymbol {\mathcal L}}}
\newcommand{\ez}{{\boldsymbol e}_z}
\newcommand{\ex}{{\boldsymbol e}_x}
\newcommand{\ey}{{\boldsymbol e}_y}
\newcommand{\ii}{{i\mkern1mu}}

\shorttitle{Imaging near-surface flows using mode-coupling analysis}
\shortauthors{Mani et al.}

\begin{document}

\title{Imaging the Sun's near-surface flows using mode-coupling analysis}

\author[0000-0002-8707-201X]{Prasad Mani}
\affiliation{Department of Astronomy and Astrophysics, Tata Institute of Fundamental Research, Mumbai, India}
\email{prasad.subramanian@tifr.res.in}

\author[0000-0003-2536-9421]{Chris S. Hanson}
\affiliation{Center for Space Science, NYUAD Institute, New York University Abu Dhabi, Abu Dhabi, UAE}

\author[0000-0003-2896-1471]{Shravan Hanasoge}
\affiliation{Department of Astronomy and Astrophysics, Tata Institute of Fundamental Research, Mumbai, India}
\affiliation{Center for Space Science, NYUAD Institute, New York University Abu Dhabi, Abu Dhabi, UAE}

\begin{abstract}
The technique of normal-mode coupling is a powerful tool with which to seismically image non-axisymmetric phenomena in the Sun. 
Here we apply mode coupling in the Cartesian approximation to probe steady, near-surface flows in the Sun. Using Doppler cubes obtained from the Helioseismic and Magnetic Imager onboard the Solar Dynamics Observatory, we perform inversions on mode-coupling measurements to show that the resulting divergence and radial vorticity maps at supergranular length scales ($\sim$30~Mm) near the surface compare extremely well with those obtained using the Local Correlation Tracking method. We find that the Pearson correlation coefficient is $\geq$ 0.9 for divergence flows, while   $\geq$ 0.8 is obtained for the radial vorticity. 

\end{abstract}


\section{Introduction} \label{sec:intro}
Helioseismology is the study of the Sun's internal structure and its properties, by means of interpreting its effect on solar oscillations \citep[see][for a review]{CD02}. These are resonant normal modes of the Sun, behaving as standing waves in a cavity bounded by the solar surface and a depth that depends on the wavenumber of the oscillation. As these waves penetrate the interior, they register information of the properties and dynamics of the solar interior and return to the surface, where they are observed. The internal structure of the Sun can then be retrieved through meticulous inversions of these seismic measurements.

Several important flow systems on the Sun have been inferred using various global and local helioseismic methods. Of those, the most notable global helioseismic results include inferences on the solar differential rotation, through global mode frequency splitting \citep[][]{Thompson96, Schou98}, and the resolving the neutrino problem \citep[][]{Bahcall92}. Notable local helioseismic results include imaging of the meridional flow \citep[][]{Giles97,Gizon20} through time-distance helioseismology \citep[][]{Duvall93}, and farside imaging of active regions \citep[][]{braun_lindsey_2001} and their near side emergence \citep[][]{Birch16}, through helioseismic holography \citep{Lindsey00}. The recent discovery of various inertial waves \citep[][]{Gizon21}, including the equatorial Rossby wave \citep{Loptein18}, has been achieved through local helioseismic ring-diagram analysis \citep{Hill88} and the non-helioseismic local correlation tracking \citep[LCT,][]{November88} of granulation. 

In recent years, the use of global mode-coupling helioseismology \citep[][]{Woodard89,Lavely92} has received attention, with many studies seeking to validate and demonstrate the importance of such a technique for investigating numerous solar phenomena. While the derivation of the mode-coupling technique is mathematically challenging, the data analysis is simple and utilizes all the information registered by the mode. Thus far,  global mode-coupling has been validated through observations of the meridional flow \citep[][]{Vorontsov11, Woodard13}, 
differential rotation \cite[][]{Schad20,Kashyap21}, global-scale convection \citep[][]{Woodard14, Woodard16, Hanasoge20, Mani21} and Rossby modes \citep[][]{Hanasoge19, Mandal20, Mandal21}. Local mode-coupling analysis in the Cartesian approximation, formulated by \citet{Woodard06}, was validated by \citet{Hanson21} (hereafter H21) by examining the power-spectrum of supergranular waves and comparing with previous time-distance studies \citep{Langfellner18}.

Normal-mode coupling refers to the concept of expressing solar-oscillation eigenfunctions as a linear weighted combination of model-eigenfunctions \citep[e.g., Model S][]{CD21}. The model eigenfunctions form a complete and orthogonal basis. By design, the model Sun is spherically symmetric, adiabatic, free from rotation, magnetism and flows. In this state, the oscillations are considered to be uncoupled. The weights needed to express the solar-oscillation eigenfunctions would then encode all the perturbations that are absent in the model. The forward problem then reduces to relating observed seismic measurements to the perturbations that we want to infer. The surface wavefield cross-correlation is the primary measurement in the mode-coupling analysis and can be directly related to the weights \citep{Woodard16}. As mode coupling is a Fourier domain technique, wavefields are cross-correlated at different spatial and temporal frequencies, leaving us with measurements sensitive to different quantities of interest. 

In this study, we extend the spectral analysis of H21 and develop the method to produce near-surface flow maps at supergranulation length scales. A part of the formalism that was used to derive the forward model in H21 is reworked, primarily to image steady flows. Measurements are then constructed, and inversions to infer divergence flow and radial vorticity are described. We also demonstrate signal associated with supergranular flow in a radial-order coupling (p$_2$-p$_2$), which was not shown in H21. This helps in localizing the measurement sensitivity to the surface. We compare our results with flows obtained using the Local Correlation Tracking method on solar granules. 

\subsection{Forward problem}\label{forward problem section}
In favor of algebraic brevity, we only show crucial steps here and refer the interested reader to Appendix~\ref{appendix a} for a complete derivation of the forward problem. Working in the plane-parallel atmosphere \citep[see also][]{Woodard06}, we denote the horizontal unit vectors $\ex$ and $\ey$ in our local Cartesian domain as pointing towards west and north on the solar surface, respectively, and $\ez$ points outwards. This approximation is valid when observing patches of the surface that are small when compared to the solar radius. When imaging steady, near-surface flows in the neighbourhood of the supergranular scale ($\sim30$~Mm), we expect the measured spectral cross-correlation signal to peak around the horizontal wavenumber $qR_\odot \approx 120$ \citep[][]{Rincon18}, where $q = |\bq| = |(q_x,q_y)|$ is the vector horizontal wavenumber of the flow. Accordingly, the goal is to relate measurements (linearly, to facilitate inversion) to the flow perturbation described in a horizontal Fourier domain. Supergranular velocities are subsonic \citep[300-400 m/s, see][]{Rincon18}, permitting us to model the flow vector $\pmb{u}=(u_x,u_y,u_z)$ in the Cartesian domain like so \citep[][]{Unno89, Woodard06}
\begin{align}\label{flow1}
    \bfu^\sigma = \bnabla\bcurl[\bnabla\bcurl(\pol\,\ez)] + \bnabla\bcurl(\tor\,\ez),
\end{align}
where $\pol = \pol^\sigma(\bx)$ and $\tor = \tor^\sigma(\bx)$ are poloidal and toroidal scalar functions, varying with position $\bx$ and temporal frequency $\sigma$. $\bnabla$ is the 3D gradient operator. While mode-coupling can easily be extended to study time-varying perturbations \citep[see][for example]{Woodard16,Mani20,Hanasoge20,Mandal20}, here we only consider the frequency bin $\sigma=0$, denoting the temporally averaged flow over the period of analysis. We therefore suppress $\sigma$ from all terms this point forward, remembering that temporal dynamics of perturbations may also be studied using the same model outlined in the following paragraphs. Simplifying eq~\ref{flow1} using vector calculus results in 
\begin{equation}\label{flow2}
    \bfu = - \bnabla^2 \pol \ez + \bnabla(\partial_z \pol) + \bnabla_h\tor \bcurl \ez,
\end{equation}
where $\bnabla_h$ refers to derivatives only in the horizontal direction. Mode-coupling helioseismology is performed in the Fourier domain, and since we wish to image horizontal flows on a small patch of the surface, we describe the flow as a function of horizontal wavenumber $\bq$ and depth $z\ez$. Hence the poloidal and toroidal flows are described by $\pol_{\bq}(z)$ and $\tor_{\bq}(z)$, respectively. Furthermore, we parametrize the flow along $\ez$ using basis functions $f(z)$ (Chebyshev, $B$-spline, etc). This is expressed as
\begin{eqnarray}\label{z-basis}
\pol \equiv \pol_{\bq}(z) = \sum_j\,f_j(z)\,\pol_{\bq j}, \,\,\,\,\,\,\, \tor \equiv \tor_{\bq}(z) = \sum_j\,f_j(z)\,\tor_{\bq j}.
\end{eqnarray}
The flow coefficients $\pol_{\bq j}$ and $\tor_{\bq j}$, represented by the discrete indices $\bq$ and $j$, become ideal candidates for inversions, where the flow for each wavenumber $\bq$ can be inverted for independently; parallelization in computation can thus be exploited to expedite inversions. Note that $\pol_{\bq j} = \pol_{-\bq j}^*$ and $\tor_{\bq j}=\tor_{-\bq j}^*$ for the flow field to be real in the spatio-temporal domain.
\\
To infer flows from wavefields $\phi$ scattered by a perturbation of length scale $\bq$, cross-correlate them in the manner $\phi^{\omega*}_\bk\,\phi^{\omega}_{\bk+\bq}$, where $\bk$ is the oscillation mode wavenumber $(k_x, k_y)$ and $\omega$ is the temporal frequency. Relate $\phi^{\omega*}_\bk\,\phi^{\omega}_{\bk+\bq}$ thus to the flow coefficients $\pol_{\bq j}$ and $\tor_{\bq j}$ (see eq~\ref{cross correlation forward integral})
\begin{equation}\label{cross-correlation forward model}
    \langle \phi^{\omega*}_\bk\,\phi^{\omega}_{\bk+\bq}\rangle = {\mathcal H}^\omega_{kk'nn'}\,\sum_j C_{\bq j,\bk} \pol_{\bq j} + \mathcal{D}_{\bq j,\bk} \tor_{\bq j}.
\end{equation}
The weight factor $\mathcal{H}^\omega$ (see eq~\ref{hfactor}) is a function of frequency, capturing information about the extent of coupling between the two modes $[n,k]$ and $[n',k']$, where $n$ and $n'$ are the radial orders of the modes, and $k = |\bk|$ and $k'= |\bk'| = |\bk + \bq|$. The spectral profile of the mode (see eq~\ref{lorentzian}) is approximated using a Lorentzian \citep[][]{Anderson90}. The more the Lorentzians of the two modes overlap, the stronger the coupling. Finally, the real terms $C_{\bq j,\bk}$ and $D_{\bq j,\bk}$ are poloidal and toroidal flow sensitivity kernels respectively, that allow us to relate the flows in question to the measurements and are derived from the solar model see Appendix~\ref{appendix a}. They possess the symmetry relation: $C_{\bq j,\bk} = C_{-\bq j,-\bk}$ and $\mathcal{D}_{\bq j,\bk}=\mathcal{D}_{-\bq j,-\bk}$ (see eq~\ref{kernels}). The kernels, as flows, are expressed on the basis $f_j(z)$.

\subsection{Least-squares of cross-correlation}
Even though $\phi^{\omega*}_\bk\,\phi^{\omega}_{\bk+\bq}$ isolates the effect of flow perturbations at individual wavenumbers $\bq$, a more compact measurement, known in mode-coupling literature as '$B$-coefficients', is much better designed for inversion as it reduces the dimension of the problem. A least-squares fit to the cross-correlation $\phi^{\omega*}_\bk\,\phi^{\omega}_{\bk+\bq}$ \citep[see][]{Woodard06, Woodard14, Woodard16} results in the $B$-coefficients $B_{\bk,\bq}$, according to 
\begin{equation}\label{least-squares}
    B_{\bk,\bq} = \frac{\sum\limits_{\omega} {\mathcal H}^{\omega*}_{kk'nn'}\phi^{\omega*}_\bk\,\phi^{\omega}_{\bk+\bq}}{\sum\limits_{\omega} |{\mathcal H}^{\omega}_{kk'nn'}|^2}.
\end{equation}
Multiplying eq~\ref{cross-correlation forward model} on both sides by ${\mathcal H}^{\omega*}_{kk'nn'}$ and substituting by eq~\ref{least-squares} on the left-hand-side results in a concisely defined forward problem (compare with eq~\ref{cross-correlation forward model})
\begin{equation}\label{forward problem}
    B_{\bk,\bq} = \sum_j C_{\bq j,\bk} \pol_{\bq j} + D_{\bq j,\bk} \tor_{\bq j}.
\end{equation}
In eq~\ref{least-squares}, \citet{Woodard07} and H21 thus far only considered positive-frequency components in the summation over $\omega$. Here, we sum over both $\pm\omega$ within a few mode linewidths $\Gamma$. Denoting the resonant frequency of a mode using $\omega_{nk}$,
\begin{align}\label{omega range}
    |\omega|\in\Big(\omega_{nk}-\epsilon\Gamma_{nk}/2,\omega_{nk}+\epsilon\Gamma_{nk}/2 \Big)\;\;\;\nonumber or \\
    |\omega|\in\Big(\omega_{n'k'}-\epsilon\Gamma_{n'k'}/2,\omega_{n'k'}+\epsilon\Gamma_{n'k'}/2\Big).
\end{align}
Summing over $\pm\omega$ guarantees that the parity $B_{\bk,\bq} = B^*_{-\bk,-\bq}$ (see Appendix~\ref{appendix a} for derivation) is obeyed, thereby ensuring that the flow field on the right-hand-side of eq~\ref{forward problem} is a real physical quantity in the spatio-temporal domain. Taking the complex conjugate on both sides of eq~\ref{forward problem} and considering the negative wavenumber components $-\bq$ and $-\bk$,
\begin{equation}
B_{-\bk,-\bq}^* = \sum_j C_{-\bq j,-\bk} \pol_{-\bq j}^* + D_{-\bq j,-\bk} \tor_{-\bq j}^*.    
\end{equation}
Substituting parity and symmetry relations for all terms in the above results in eq~\ref{forward problem}. As $B_{\bk,\bq}$ is constructed by a least-squares fitting, it is noteworthy that summing over $-\omega$ will also lead to improvement in its signal-to-noise as a by-product.

\subsection{Noise model}
In the addition to the sensitivity kernels, a systematic background noise model is required to infer the flows from the observed $B$-coefficients. For estimating the contribution from realization noise to the measurements, we make the following assumptions \citep[][]{Gizon04}: that the excitation of the wavefield is modelled as a multivariate Gaussian random process and the wavefields are uncorrelated across wavenumber and frequency in the absence of perturbations. Every independent realization of a mode can be understood as the output of a damped harmonic oscillator driven by a random forcing function \citep[see][]{Duvall86}. Modes are thus generated with random phases and amplitudes and with finite lifetimes. This stochasticity leads to realization noise in repeated measurements of mode parameters such as its amplitude, frequency and linewidth, and consequently in $B_{\bk,\bq}$ in our case. We use the same noise model as in H21, which was motivated by the above discussion,
\begin{equation}\label{noise model}
    G_{\bk,\bq}\equiv\langle|B_{\bk,\bq}|^2\rangle,
\end{equation}
where, unlike H21, we again sum over $\pm\omega$. $G_{\bk,\bq}$ is real, with the symmetry relation $G_{\bk,\bq} = G_{-\bk,-\bq}$ (see Appendix~\ref{appendix a} for explanation).

\section{Data Analysis}\label{Data Analysis}
In order to examine near surface flows we build a time-series cube of Doppler images that are obtained from the Helioseismic Magnetic Imager aboard the Solar Dynamics Observatory \citep[HMI/SDO,][]{Scherrer12}. Each image is Postel projected, with a spatial resolution of approximately $0.48$Mm, sperated in time by 45 seconds, and is tracked at the \citep[][]{Snodgrass84} rotation rate. Here, we select a patch that is $194.4\times194.4$ Mm$^2$ in size, tracked for 24 hours and crosses the disk-center in the middle of observation time on the 14 Novemeber 2017 (Carrington rotation number 2197, Carrington longitude $90^\circ$). This Dopplercube is considered as the physical wavefield $\phi(x,y;t)$. The Fourier-space wavefield $\phi^{\omega}_\bk$ (and subsequently, the cross-correlation $\phi^{\omega*}_\bk\,\phi^{\omega}_{\bk+\bq}$) is obtained by computing the 3D spatial and temporal Fourier transform of the Dopplercube.

The duration of the observed region is long enough to provide sufficient frequency bins with which to sum over in Eq~\ref{forward problem}, while short enough that supergranules do not substantially evolve \citep[lifetime is purported to be 1.6 days;][]{Rincon18} over this period. Our observation region is close to the disk center to also avoid any contamination from center-to-limb systematics \citep[][]{Zhao12,Langfellner15}.

Maximum signal can be extracted from the weighted summation of the cross correlations (eq~\ref{least-squares}) when the spectral profiles of the two modes $[n,k]$ and $[n',k']$ closely align in $\omega$ space. This implies that their mode frequencies should be sufficiently close ($|\omega_{nk} - \omega_{n'k'}|\leq\delta$, the separation parameter). Since Lorentzians decay rapidly, the summation over $\pm\omega$ is significant only over a few linewidths ($\epsilon$, the summation parameter; see eq~\ref{omega range}). We have empirically found and tabulated $\delta$ in Table~\ref{separation table} for the radial order couplings $n$-$n'\in$ f-f, p$_1$-p$_1$, and p$_2$-p$_2$ (the signal strength depends only weakly on $\epsilon$; we set it to 3 line widths).

\begin{figure}
\subfloat{\includegraphics[width = 3.2in]{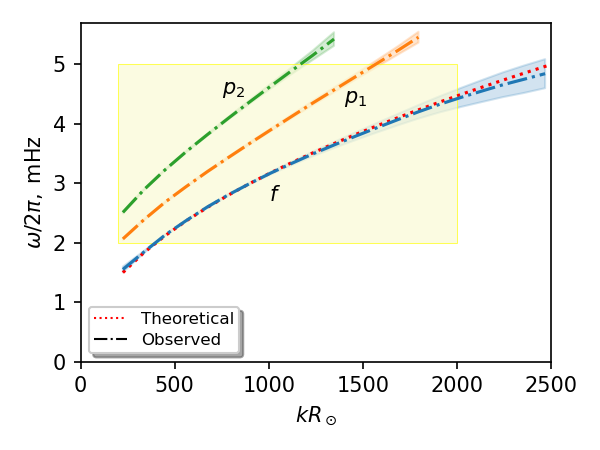}}
\caption{Dispersion relation for the radial orders used in this analysis; f (blue), p$_1$ (orange) and p$_2$ (green). The shaded regions of the same colours indicate 1-linewidth $\Gamma$ about the mode frequency. The yellow shaded region indicates the range of $kR_\odot$ and $\omega/2\pi$ to which we have restricted ourselves in this analysis. Beyond $kR_\odot$ of 2000, it is seen that the theoretical fitting of mode frequencies start deviating from the observed dispersion relation for the $f$-mode.}
\label{lnu}
\end{figure}

Figure~\ref{lnu} shows that for any two adjacent ridges (adjacent $n$ and $n'$), mode frequencies $\omega_{nk}$ and $\omega_{n'k}$ become spaced farther apart with increasing wavenumber $kR_\odot$. It is also known that mode linewidth $\Gamma$ grows with radial orders for a given $kR_\odot$. Moreover, holding the spatial and temporal sampling rates constant, the spatial size and duration of observation set the total number of modes within a range of $kR_\odot$ (and $\omega/2\pi$) that can be clearly observed, thereby affecting the quality of the seismic measurements. Owing to these factors, to maximize signal-to-noise (by visually inspecting the power-spectrum), the  parameters describing the extent of coupling over different ranges of $kR_\odot$ at fixed radial order are different. In wavenumber, we restrict our analysis to within $200\leq kR_\odot\leq2000$ and $qR_\odot\leq300$. Our frequency range is confined to span the range over which acoustic modes are observed ($2 \leq \omega/2\pi \leq 5$ in mHz). 

\begin{table}[h!]
\setlength{\arrayrulewidth}{0.4mm}
\setlength{\tabcolsep}{8pt}
\renewcommand{\arraystretch}{1.2}
\centering
\begin{tabular}{c|ccc}
\hline
\hline
Coupling & k$R_\odot$ range & \# of & $\delta$\\
& & modes&  \\
\hline
f-f & [400,1000] & 5240 & 4\\
 & [1000,1500] & 7784 & 1.1\\
 & [1500,2000] & 10940 & 0.4\\
\hline
p$_1$-p$_1$ & [400,1000]  & 5240 & 4.5 \\
 & [1000,1750] & 12852 & 2\\
\hline
p$_2$-p$_2$ & [200,1000] & 5886 & 3\\
 & [1000,1300] & 4280 & 3\\
\hline
\end{tabular}
\caption{Total number of modes, and separation parameter (in number of linewidths) for different couplings, for different ranges of $kR_\odot$.}
\label{separation table}
\end{table}

\section{Inversion}\label{inversion}
The final step to producing near-surface flow maps in Cartesian mode-coupling is to invert the measurements $B_{\bk,\bq}$ from the linear relation in eq~\ref{forward problem}. We describe inversion using regularized-least-squares (RLS) method here and leave Subtractive Optimally Localized Averages \citep[SOLA,][]{Pijpers94} for Appendix~\ref{appendix b}. The methods complement each other \citep[see][]{Takashi97}, where RLS tries to minimize the misfit between data and model, whereas SOLA gives better localization. For total number of modes $M$, RLS scales as $M$x$J$ where $J$ is the number of basis functions $f_j(z)$ ($J\ll M$; see eq~\ref{z-basis} and section~\ref{RLS}), whereas SOLA scales as $M^2$ (see Appendix~\ref{appendix b}). For $M>5000$, computation starts to quickly become expensive for SOLA. 

Mode eigenfunctions peak near the surface, with higher radial orders possessing smaller peaks in the interior. While f-f coupling alone has enough sensitivity to probe perturbations at supergranular scales close to surface, signal is present even in p$_1$-p$_1$, and p$_2$-p$_2$ (see Figure~\ref{all_ps}), and possibly other higher order self- and cross-couplings. Since we are interested in only surface flows, we leave higher order coupling to future work.

It bears mentioning that the slopes of the ridges in the $kR_\odot$-$\nu$ spectrum (Figure~\ref{lnu}) increase with radial order. This limits us to low-to-intermediate $kR_\odot$ ($<1000$) for these higher radial orders if we are to remain under the acoustic cut-off frequency of 5.3mHz. It also becomes imperative to use a spatially larger observation patch to gain access to signals from low $kR_\odot$ - too large an observation region could possibly render invalid the Cartesian geometry approximation. Regardless, in addition to performing inversions using all the couplings stacked together, we also demonstrate inversions separately for the three couplings (see Table~\ref{correlation table}) in order to account for the full gamut of mode-coupling as a signal-rich helioseismic technique.

\subsection{RLS}\label{RLS}
For given $\bq$, the forward problem may be stated as
\begin{equation}\label{eqRLS}
    \mathbf{K}\mathbf{U} = \mathbf{B},
\end{equation}
with the aim to minimize the misfit $\sum\limits_{k}\;||\mathbf{K}\mathbf{U}-\mathbf{B}||_2$, with $||\;||_2$ denoting the $L_2$ norm.
Here, $\mathbf{K}$ is the matrix formed by the sensitivity kernels: $\{C_{\bq j,\bk},\mathcal{D}_{\bq j, \bk}\}$. $\mathbf{U}$ is a vector composed of the flow coefficients: $\{\pol_{\bq j},\tor_{\bq j}\}$ and $\mathbf{B}$ is a vector composed of computed $B$-coefficients: $\{B_{\bk,\bq}\}$. The least-squares problem is solved simultaneously for poloidal and toroidal flow. We use $B$-spline basis functions as our $f_{j}(z)$, comprising 11 knots spaced uniformly in acoustic radius, for both poloidal and toroidal coefficients. 
Hence, for $M$ modes (total number of $\bk$ for a given $\bq$ is $M$) and 11 basis functions for each poloidal and toroidal, the dimensions of $\mathbf{K}$, $\mathbf{U}$ and $\mathbf{B}$ are thus $M\times22$, $22\times1$, and $M\times1$ respectively. Normalizing both sides of eq~\ref{eqRLS} by the noise covariance $\mathbf{\Lambda}$ (a diagonal matrix with the entries $G_{\bk,\bq}$; see eq~\ref{noise model}; dimension $M\times M$) and pre-multiplying by $\mathbf{K}^\intercal$,
\begin{align}
    (\mathbf{K}^\intercal\mathbf{\Lambda}^{-1} \mathbf{K})\mathbf{U} =& (\mathbf{K}^\intercal\mathbf{\Lambda}^{-1})\mathbf{B},\\
    \mathbf{U} =& ( \mathbf{K}^\intercal\mathbf{\Lambda}^{-1}\mathbf{K})^{-1} \mathbf{K}^\intercal\mathbf{\Lambda}^{-1}\mathbf{B}.
\end{align}

\begin{figure}
\subfloat{\includegraphics[width = 6.2in]{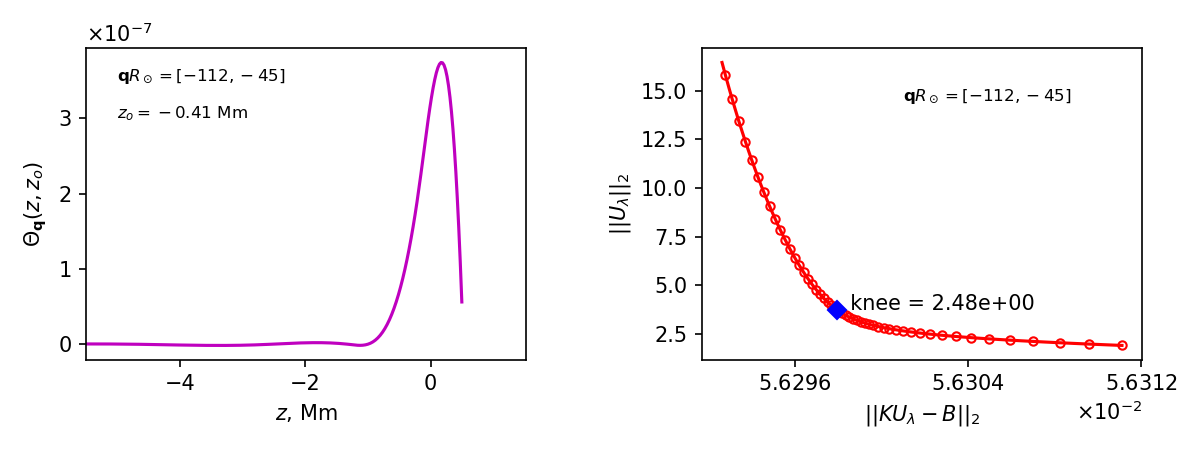}}
\caption{\textit{Left}: Averaging kernel for poloidal flow 
(see section~\ref{section alpha}, eq~\ref{avgkernel}, and left panel of Figure~\ref{fig_kernel}) for $\bq R_\odot=[-112,-45]$, at the depth $z_o=-0.41$ Mm. \textit{Right}: L-curve for the mode $\bq R_\odot=[-112,-45]$; the knee ($\lambda=2.48$) is marked by a blue diamond.}
\label{rlslcurve}
\end{figure}
\begin{figure}[!htb]
\subfloat{\includegraphics[width = 6.5in]{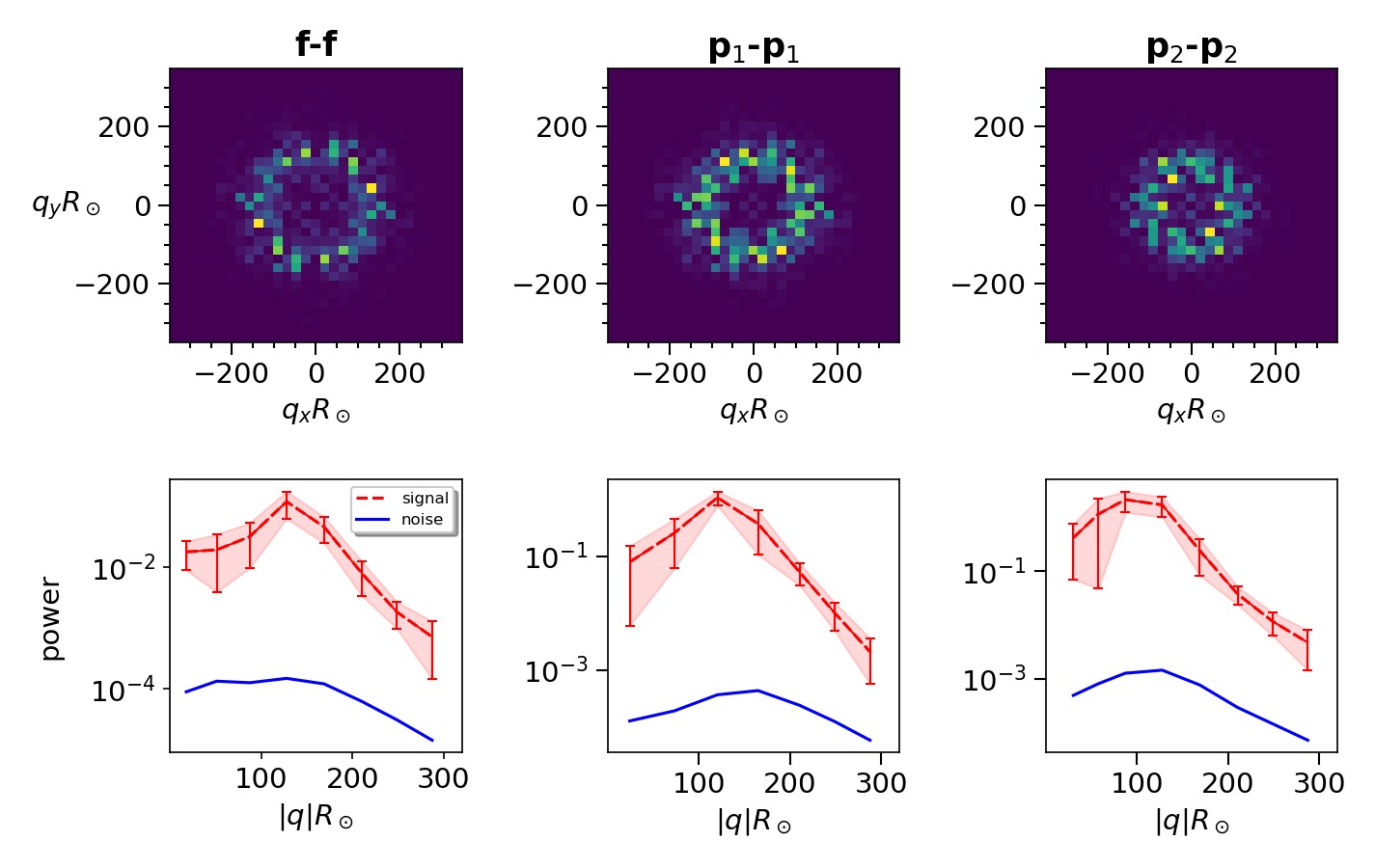}}
\caption{\textit{Top}: Inverted poloidal flow power-spectrum for the three couplings f-f, p$_1$-p$_1$, and p$_2$-p$_2$ as a function of $q_x R_\odot$ and $q_y R_\odot$. \textit{Bottom}: Corresponding power-spectrum averaged over the azimuthal angle. Shaded region shows $\pm1\sigma$ error around the mean. Total power appears to increase through the radial orders. Power is in units of m$^2$/s$^4$.}
\label{all_ps}
\end{figure}

Since the least-squares problem is typically ill-posed, we restate the minimization as $\sum\limits_{\bk}\;||\mathbf{K}\mathbf{U}-\mathbf{B}||_2 + \lambda||\mathbf{U}||_2$ with the regularization parameter $\lambda$ which this results in a trade-off between misfit reduction (first term) and solution norm minimization (second term). Under-regularizing can lead to a solution $\mathbf{U}$ that is dominated by errors in the data and on the other hand, over-regularizing may smooth or damp the solution more than necessary. Including this regularization makes the problem better conditioned and is now defined as 
\begin{eqnarray}\label{rlsinversion}
           \mathbf{U} = ( \mathbf{K}^\intercal\mathbf{\Lambda}^{-1}\mathbf{K} + \lambda\mathbf{I})^{-1} \mathbf{K}^\intercal\mathbf{\Lambda}^{-1}\mathbf{B},
\end{eqnarray}
where $\mathbf{I}$ is the identity matrix for $L_1$ regularization. The knee-point of the $L$-curve \citep{hansen_1992}, a curve formed by plotting $||\mathbf{U}||_2$ vs $||\mathbf{K}\mathbf{U}-\mathbf{B}||_2$ for different values of $\lambda$ (see right panel of Figure~\ref{rlslcurve}),  is usually chosen as the regularization parameter. After successfully inverting for $\mathbf{U}$, we reconstruct the flow using eq~\ref{z-basis}. Results for poloidal flow $\pol_{\bq}$ are shown in Figure~\ref{all_ps}. 

\section{LCT}
To improve confidence in the imaged near-surface flows through mode-coupling, we compare them with flows obtained from Local Correlation Tracking method \citep[LCT;][]{November88}. LCT provides surface-flow maps by examining the advection of convective granules \citep[1.2 Mm, $qR_\odot\approx3500$;][]{Hathaway15} by underlying larger-scale flow systems. Since granules are used as tracers, which are much smaller in size than supergranules ($\approx35$ Mm), LCT is an effective method  \citep[see][]{Rieutord01} to produce surface horizontal flow maps of supergranulation.

Time series of intensity images from HMI, with the same properties of the Dopplercubes described in section~\ref{Data Analysis} (tracking rate, date, location, size and duration of observed patch, spatial and temporal sampling rate), are obtained and Postel projected. The horizontal flows are deduced by tracking the proper motions of granules between consecutive intensity images, which we denote as $\mathcal{I}_1,\mathcal{I}_2$. The LCT method selects a patch in two images each ($I_1 = \mathcal{I}_1e^{(\bx-\bx_{ij})^2/2\,\mathtt{sigma}^2}, I_2 = \mathcal{I}_2e^{(\bx-\bx_{ij})^2/2\,\mathtt{sigma}^2}$) that observe the same granule at the grid point $\bx_{ij} = (x_i,y_j)$. A Gaussian of width $\mathtt{sigma}$ allows to isolate a small region surrounding the grid point of interest as the distance moved by granules are usually in sub-pixel regime. The convention for the direction of $\bx$ is the same as described in section~\ref{forward problem section}. The two patches $I_1,I_2$ are then cross correlated for different values of position shifts $\Delta \bx$, 
\begin{equation}
    C_{ij}(\Delta x, \Delta y) = \int \textrm{d}\bx\; I_1^*(-\bx)I_2(\Delta \bx -\bx).
\end{equation}
The shift $\Delta \bx = (\Delta x, \Delta y)$ that maximizes the cross-correlation $C_{ij}$ is taken to be the proper motion of the granule. Provided that the time difference $\Delta t$, here 45 seconds, between the images is less than the lifetime of granules ($<10$ min), the velocities are given by $v_x = \Delta x / \Delta t$ and $v_y = \Delta y / \Delta t$. This exercise is repeated for all grid points in the images $\mathcal{I}_1,\mathcal{I}_2$ and for each consecutive pair of images in the cube. 

In practice, we use the Fourier LCT algorithm \citep[FLCT,][]{FLCT} for computing $v_x$ and $v_y$. FLCT requires the input $\mathtt{sigma}$, which we set to 4 pix, that captures the extent of localization desired, and depends on the dominant length scale of the velocity field in the images. The Postel-projected intensity images are fed as input to the FLCT code. $v_x$ and $v_y$ are then computed for consecutive pairs of images and are averaged over the entire day.  

\section{Maps of horizontal divergence and radial vorticity}
For mode-coupling, horizontal divergence (hereafter $div$) and radial vorticity (hereafter $curl$) are computed by substituting $\pol$ and $\tor$ from eq~\ref{z-basis} into eq~\ref{flow2} as below - 
\begin{align}
    \pmb{u}(\bq,z)&= - \bnabla^2 \pol\ez + \bnabla(\partial_z \pol) + \bnabla_h\tor \bcurl \ez\nonumber,\\
     &=- (0,\,0,\,\partial_x^2\pol+\partial^2_y\pol+{\partial^2_z \pol}) + (\partial_x\partial_z\pol, \, \partial_y\partial_z\pol, \, {\partial^2_z \pol}) + (\partial_y\tor, \, -\partial_x\tor, \, 0).
\end{align}
Setting $\partial_x ^2 + \partial_y ^2 = q^2$, $div$ is given by, 
\begin{align}\label{mcdiv}
    \bnabla_h\cdot\pmb{u}(\bq,z) &= q^2\partial_z \pol,
\end{align}
and $curl$ is given by, 
\begin{align}\label{mccurl}
    \Big[\bnabla \times \pmb{u}(\bq,z)\Big]_z &= q^2\tor.
\end{align}

\begin{figure}
\subfloat{\includegraphics[width = 7.2in]{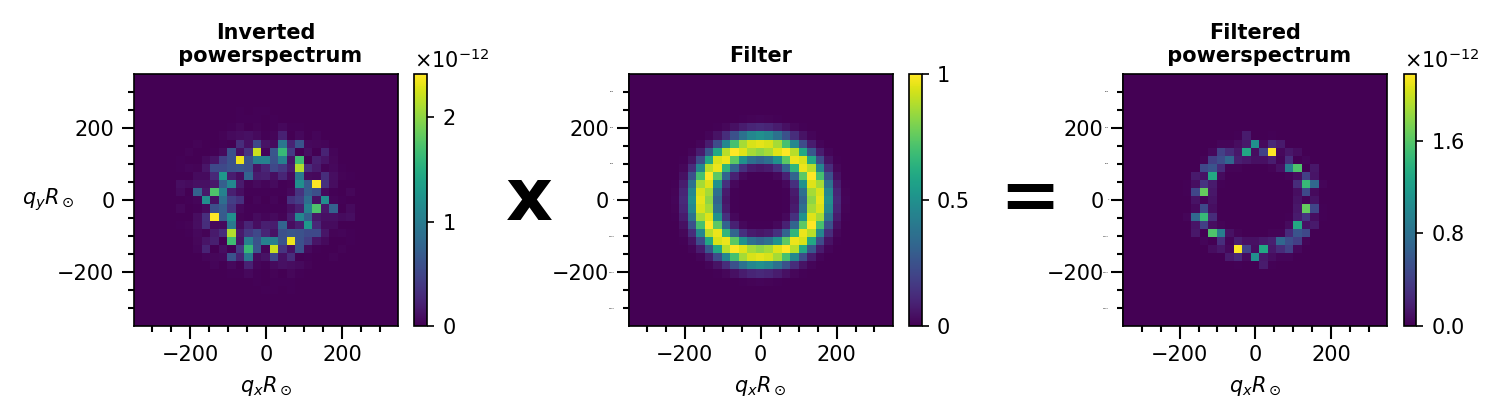}}
\caption{Left: Divergence-flow power spectrum $|div|^2$, from eqn~\ref{mcdiv}, obtained from inversion using all the couplings. The power-spectrum is then filtered with a bandpass centered around $qR_\odot=150$ (middle panel). The resulting spectra is shown in the right panel. The units of $|div|^2$ are in s$^{-2}$. For illustration, we show the action of the filter on the power-spectrum $|div|^2$ since it is a real quantity, but recall that it is the Fourier-space flow $div$ (a complex quantity) on which we apply the filter.}
\label{rls_ps}
\end{figure}

We follow similar steps to those taken in \citet{Langfellner15} for comparison of flow maps with LCT. The essential step for comparison at different length scales is to bandpass filter the Fourier-space flow around the $qR_\odot$ of interest (see Figure~\ref{rls_ps}), and subsequently convert it to real space.

\begin{figure}
\gridline{\fig{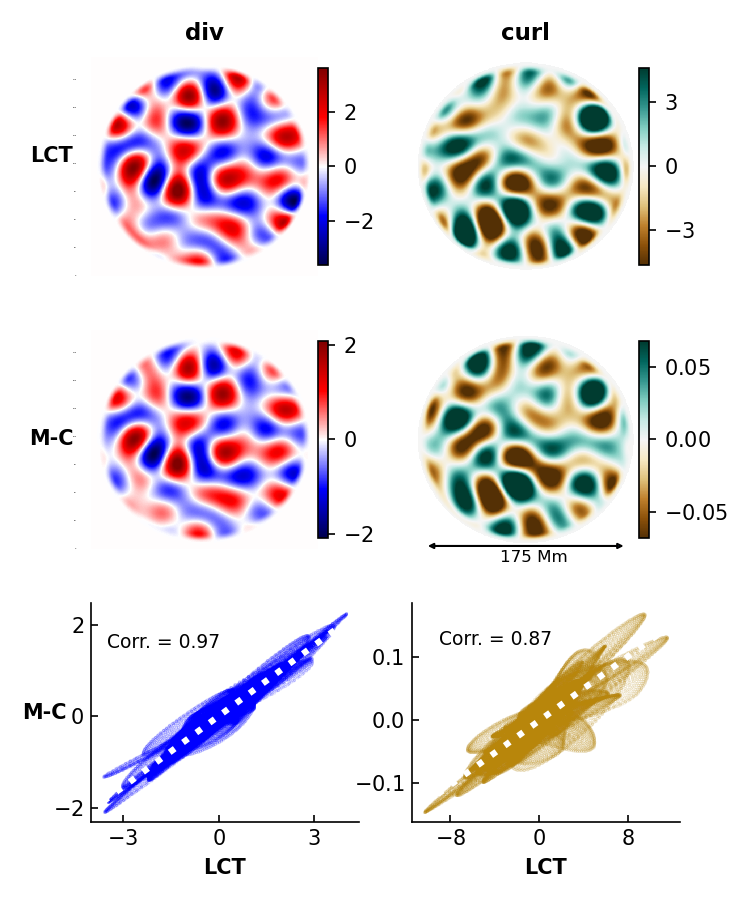}{0.5\textwidth}{(a) $qR_\odot=100,$ f-f + p$_1$-p$_1$ + p$_2$-p$_2$}}
\caption{Real-space divergence flows (left column, in units of $10^{-5}$s$^{-1}$) and radial vorticity (right column, in units of $10^{-6}$s$^{-1}$) for LCT (top row), and mode-coupling inversions through RLS using all the couplings (middle row), bandpass filtered around $qR_\odot=100$ (see Figure~\ref{rls_ps}). Corresponding scatter plots and correlation coefficients are shown in the bottom row. We cut edges out from the flow maps and compare a circular region of diameter $\approx$175 Mm. The slopes of the best-fit line through the scatter plots are 0.51 for divergence and 0.01 for vorticity. The vorticity flow maps are saturated to show only 40\% of the maximum values.}
\label{real_space_all}
\end{figure}

\begin{figure}
\gridline{\fig{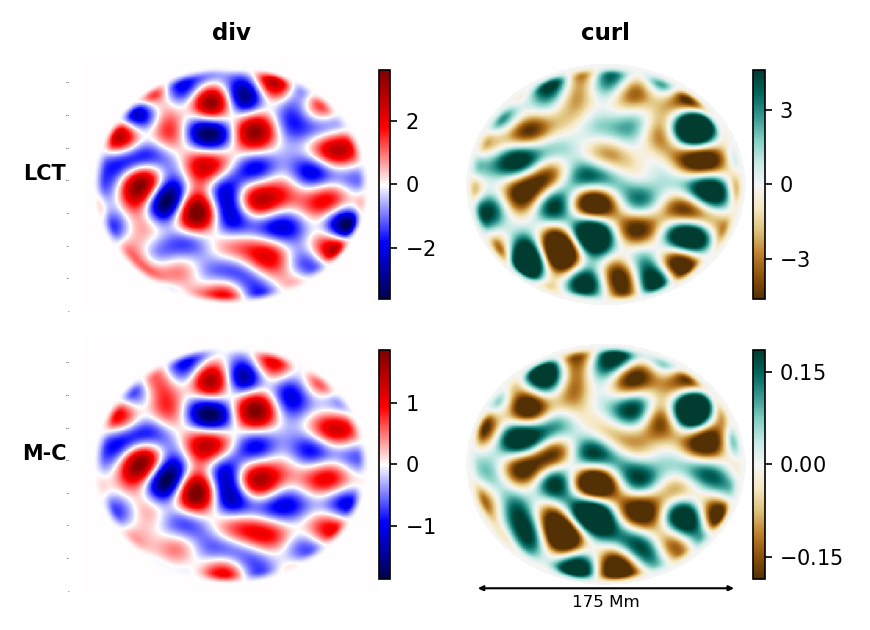}{0.5\textwidth}{(a) $qR_\odot=100,$ f-f}
\fig{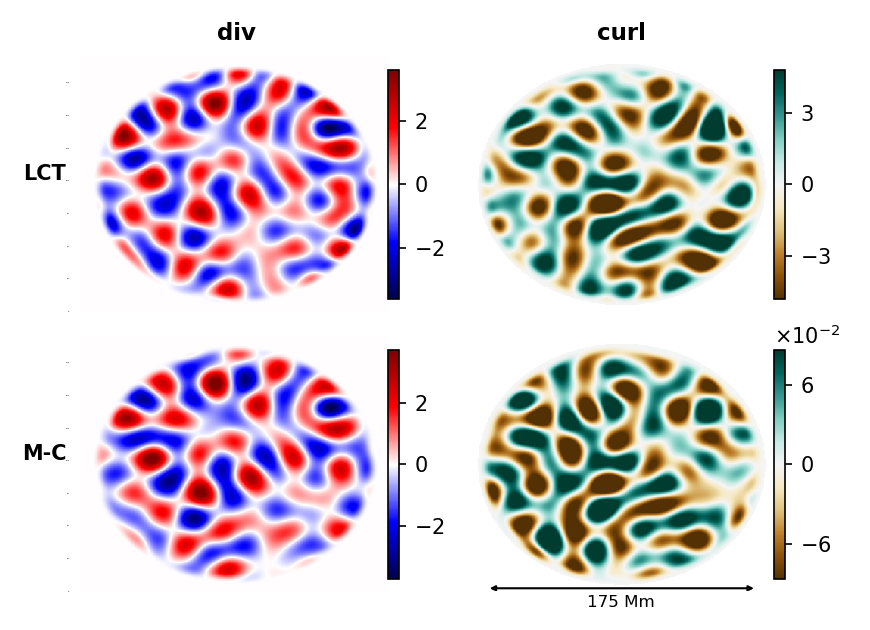}{0.5\textwidth}{(b) $qR_\odot=150,$ p$_1$-p$_1$}}
\caption{Real-space divergence flows (left column, in units of $10^{-5}$s$^{-1}$) and radial vorticity (right column, in units of $10^{-6}$s$^{-1}$) for LCT (top row), and  mode-coupling inversion through RLS using (a) f-f coupling (bottom row), bandpass filtered around $qR_\odot=100$, and using (b) p$_1$-p$_1$ coupling (bottom row), bandpass filtered around $qR_\odot=150$. We cut  edges out from the flow maps and compare a circular region of diameter $\approx$175 Mm.}
\label{real_space_100_150}
\end{figure}

\begin{figure}
\gridline{\fig{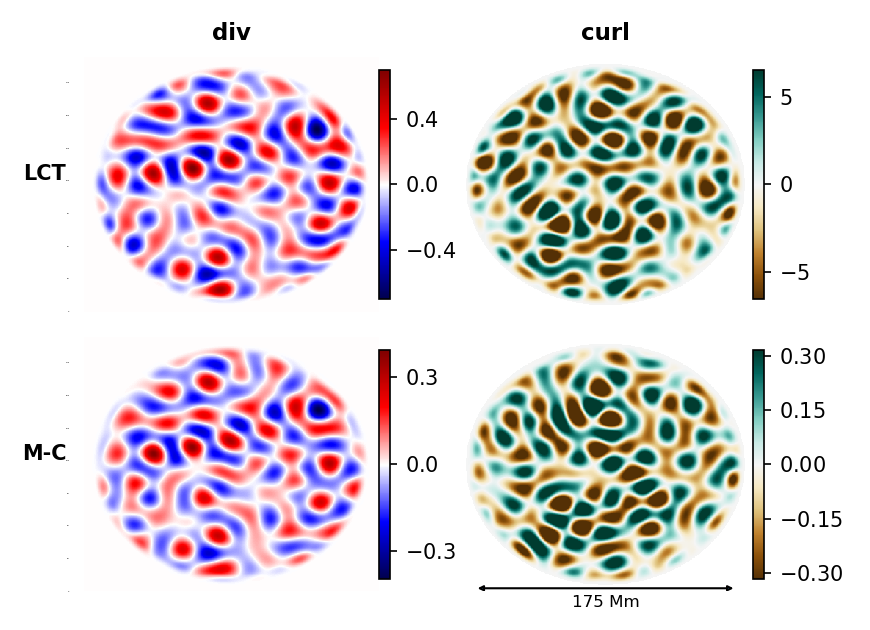}{0.5\textwidth}{(a) $qR_\odot=200,$ f-f + p$_1$-p$_1$ + p$_2$-p$_2$}
\fig{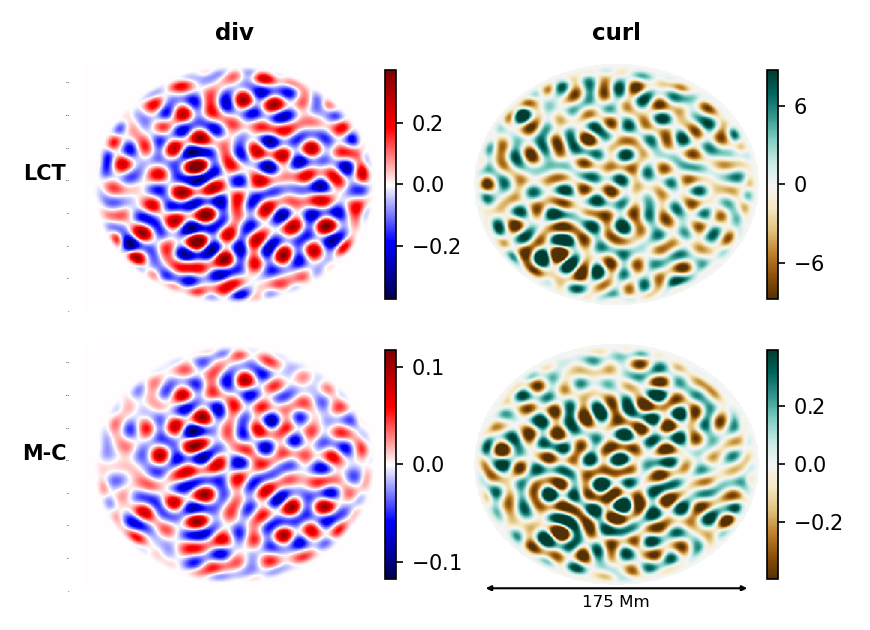}{0.5\textwidth}{(b) $qR_\odot=250,$ f-f + p$_1$-p$_1$ + p$_2$-p$_2$}}
\caption{Real-space divergence flows (left column, in units of $10^{-5}$s$^{-1}$) and radial vorticity (right column, in units of $10^{-6}$s$^{-1}$) for LCT (top row), and  mode-coupling inversion through RLS using all the couplings (bottom row), bandpass filtered around (a) $qR_\odot=200$, and (b) $qR_\odot=250$. We cut edges out from the flow maps and compare a circular region of diameter $\approx$175 Mm.}
\label{real_space_200_250}
\end{figure}

We seek to show comparisons (see Figures~\ref{real_space_all},~\ref{real_space_100_150}, and~\ref{real_space_200_250}) for $qR_\odot  = 100$, $150$, $200$ and $250$. To sufficiently delineate flows at these length scales, we apply a Gaussian filter (see Figure~\ref{rls_ps}) to flows obtained from eqns~\ref{mcdiv} and~\ref{mccurl}. The Gaussian is centered at the desired wavenumber with a half-width of 25. We then perform a 2D Fourier transform to obtain a real-space steady-flow map.

For LCT, we first apply a Gaussian smoothing to $v_x$ and $v_y$ to average over small-scale features; the extent of smoothing depends on the length scale $qR_\odot$ to be compared with mode-coupling. $div$ and $curl$ are then simply computed by 
\begin{align}
    div &= \partial_x v_x + \partial_y v_y,\label{lctdiv} \\
    curl &= \partial_x v_y - \partial_y v_x.\label{lctcurl}
\end{align}
We then perform a 2D Fourier transform on eqns~\ref{lctdiv} and~\ref{lctcurl}, apply the same Gaussian filters as for mode-coupling, and transform back to real space. 

Condensing all of the above, the following sequence of operations to compare flows at desired length scales are performed for mode-coupling (M-C) and for LCT - 
$$\text{M-C}:\;\; \phi(x,y;t)\xRightarrow[\text{}]{\text{3D FFT}}\phi_{\bk}^\omega,B_{\bk,\bq}\xRightarrow[]{\text{inversion}} \pol,\tor\xRightarrow[\bnabla \times]{\bnabla_h \cdot}\text{eqns}~\ref{mcdiv},~\ref{mccurl}\xRightarrow[\text{2D FFT}]{\text{Filter,}} div,curl$$

$$\text{LCT}:\;\;\mathcal{I}_1,\mathcal{I}_2\xRightarrow[]{\text{FLCT}} v_x,v_y\xRightarrow[\bnabla_h \cdot \;\;\bnabla \times]{\text{smooth,}}\text{eqns}~\ref{lctdiv},~\ref{lctcurl}\xRightarrow[\text{Filter}]{\text{2D FFT,}}\parbox[c]{5.5em}{\small{Filtered,  Fourier-space flows\hfil}}\xRightarrow[]{\text{2D FFT}}div,curl$$

\begin{table}[h!]
\setlength{\arrayrulewidth}{0.4mm}
\setlength{\tabcolsep}{15pt}
\renewcommand{\arraystretch}{1.2}
\centering
\begin{tabular}{c|ccc}
\hline
\hline
Coupling & q$R_\odot$ & $div$ & $curl$ \\
\hline
f-f & 100 & 0.97 & 0.87\\
+ p$_1$-p$_1$ & 150 & 0.95 & 0.76  \\
+ p$_2$-p$_2$ & 200 & 0.92 & 0.76 \\
 & 250 & 0.85 & 0.65\\
\hline
f-f & 100 & 0.96 & 0.85 \\
& 150 & 0.93 & 0.76 \\
& 200 & 0.89 & 0.69\\
& 250 & 0.77 & 0.58\\
\hline
p$_1$-p$_1$ & 100 & 0.95 & 0.83\\
& 150 & 0.95 & 0.75\\
& 200 & 0.92 & 0.75\\
& 250 & 0.85 & 0.61\\
\hline
p$_2$-p$_2$ & 100 & 0.94 & 0.7\\
& 150 & 0.91 & 0.39\\
& 200 & 0.79 & 0.3\\
& 250 & 0.55 & 0.3\\
\hline
\end{tabular}
\caption{Correlation between mode-coupling flow maps and LCT maps derived from HMI Dopplergrams and intensity images, respectively.}
\label{correlation table}
\end{table}

\section{Results}
Table~\ref{correlation table} summarizes the results of the comparison between flows obtained from mode-coupling and LCT. Figure~\ref{real_space_all}, where we have used all the couplings to perform inversions, shows a 97\% correlation between divergence flows from the two methods near supergranular scale ($qR_\odot\approx100$). Near-surface flows are imaged most faithfully when all the couplings are used. Since vortical flows are imaged at a region near the equator, it is possible that the source of vorticity is something other than Coriolis force. Nevertheless, there is also a very good agreement (87\%) between the vortical flows as inferred from the two methods, despite being an order of magnitude weaker than the divergence flows \citep[this is consistent with the results of][]{Hathaway15,Langfellner15,Rieutord17}. Due to insufficient modes for the p$_2$-p$_2$ case (see Table~\ref{separation table}), we are unable to infer vortical flows with conviction other than near the supergranular scale, as can be seen from  Table~\ref{correlation table}. Figure~\ref{real_space_100_150} also aligns with what we believe can be accomplished through mode-coupling helioseismology - using f-f or p$_1$-p$_1$ alone to seismically infer near-surface divergence and vortical flows at different scales ($qR_\odot=100,150$) can yield extremely good agreement with LCT. As the length scale of the inferred flow moves further away from that of supergranules (Figure~\ref{real_space_200_250}), the demand on signal-to-noise also increases. An adequate number of modes (and coupling strength between higher radial-orders) thus becomes a necessity to comment substantively on the flows at these scales. 

\subsection{Amplitudes of mode-coupling flows}
For both LCT and mode-coupling divergence and vorticity maps, numerous factors, arising from the associated numerous data processing steps, can influence the final inference of flow amplitudes, making it difficult to put forward a precise statement on them. H21 reported a 60\% greater amplitude for p$_1$-p$_1$ over f-f coupling (Figure~\ref{all_ps} reflects a similar conclusion), another element to consider when combining different radial orders. The choice of regularization (see right panel of Figure~\ref{rlslcurve}) has the potential to affect the amplitudes of the inverted flows to some degree. Flow amplitudes also vary with depth, implying that different radial orders and LCT will measure different flow averages. This variability emerges as a natural consequence of any helioseismic inversion procedure necessitating the use of a radial grid along which kernels and flows tend to be described.

Thus, the amplitudes of the mode-coupling flows (and the correlation coefficient) depend upon the following factors:
\begin{itemize}
    \item Coupling(s) used,
    \item Regularization parameter in the inversion,
    \item Smoothing applied to LCT flows (indirectly; see below paragraph),
    \item The depth at which flows are inferred.
\end{itemize}

Here, we report in Table~\ref{correlation table} only the maximum correlation found from among the points in the radial grid close to the surface (within $\pm$0.5 Mm from $z$=0). For a desired comparison length scale $qR_\odot$, we first fix the coupling(s) and the regularization parameter to be used in the inversion. We then separately compute filtered divergence and vorticity maps for LCT for different values of smoothing. These flow maps are then compared with those obtained from inversions at all depths in the radial grid that are within 0.5 Mm from the surface. The highest correlation (corresponding to the above depths and smoothing) is noted and comparison flow maps are plotted for the desired $qR_\odot$.

It has been shown \citep[see][]{DeRosa04,Langfellner15} that line-of-sight velocity from Dopplergrams and LCT agree closely in amplitudes. But, to recapitulate, a host of factors described above can skew the amplitudes for divergence flows owing to the multi-step process involved in obtaining them. For example, there has been a history \citep[see, e.g.,][]{DeRosa00,Sekii07,Zhao07,Langfellner18,Vincent20,Korda21} of using travel-time difference as only a proxy for horizontal divergence. However, \citet{Langfellner15}, \citet{Birch16} and \citet{Birch19} use empirically determined conversion factors to align flow amplitudes from travel-time measurements with those of LCT, while acknowledging that LCT underestimates magnitudes \citep[see][]{Verma13, Loptein16}. Even for the case of supergranulation divergence maps obtained through ring-diagram helioseismology, \citet{Greer16} only report normalized amplitudes.

In this work, we have developed inversions to show that the Cartesian approximation of mode-coupling can be used with great confidence to investigate flows near the surface. Careful inversions of mode-coupling measurements, built using a sufficiently large modeset that penetrates into the deeper layers of the convection zone, can also enable probing of the depth structure and time-evolution of supergranules, part of future work. With enough modes to improve signal-to-noise through larger observation sizes, we suggest that Cartesian mode-coupling can find local helioseismic applications to investigate other depth- and time-varying features such as giant cell flows \citep[see][]{Hathaway13,Hanson20}, emerging active regions, meridional flows and Rossby waves.

\appendix
\section{Derivation of the Forward Model}\label{appendix a}
As described in section~\ref{forward problem section}, we seek to describe the flow $\bfu$ as a function of $\bq$ along $\ez$. To that end, substituting eq~\ref{z-basis} into eq~\ref{flow2}, 
\begin{equation}\label{expand z-basis}
\bfu^\sigma_{\bq}(z) = \sum_j\left\{q^2\,f_j \ez + i\bq\,f'_j\right\}\,\pol^\sigma_{j\bq} + i\bq\bcurl\ez\,f_j \tor^\sigma_{j\bq}. 
\end{equation}
For flows in the anelastic limit ($\bfu\ll$ speed of sound), we can denote the flow perturbation operator as $\delta\bL^\sigma = -2i\omega\rho\bfu^\sigma\cdot\bnabla$ \citep[see][]{Hanasoge17}. Substituting Eq.~\ref{expand z-basis} into the operator, we get,
\begin{eqnarray}
    \delta\bL^\sigma_\bq =& -2\ii\omega\,\rho\,(\ii\bfu_\bq^\sigma\cdot\bk + \bfu^\sigma_\bq\cdot\ez\partial_z),\\
    =&-2\ii\omega\rho\sum\limits_j \left\{- \bk\cdot\bq\,f'_j\pol^\sigma_{j\bq} - \bk\cdot(\bq\bcurl\ez)\,f_j \tor^\sigma_{j\bq}+q^2\,f_j\pol^\sigma_{j\bq}\,\partial_z \right\}.
\end{eqnarray}
Express the mode eigenfunction describing oscillations in the Cartesian domain by \citep[see][]{Woodard06}
\begin{equation}
   \bxi_{k} \equiv \bxi_{nk}(z) = \ii \hat{\bk} H_{nk}(z)\ez  + \hat{z}V_{nk}(z),
\end{equation} 
where $H$ and $V$ are real-valued functions; $n$ and $n'$ are dropped for compactness of notation. Then the coupling of two modes $\bxi_k$ and $\bxi_{k'}$ ($\bk' = \bk+\bq$), by the flow perturbation operator $\delta\bL^\sigma_\bq$, denoted by coupling integral $\Lambda_{\bk'}^{\bk}(\sigma$), is given by 
\begin{align}\label{coupling integral}
     \Lambda_{\bk'}^{\bk}(\sigma) \equiv \int \textrm{d}\bx \,(\delta\bL^\sigma_\bq\bxi_k) \cdot\bxi^*_{k'}  =& \int \textrm{d}\bx\, \Bigg[
    -2\ii\omega\rho\sum\limits_j\left\{ q^2\,f_j\pol^\sigma_{j\bq}\,(\hat{\bk}\cdot\hat{\bk}'\,H'_k H^*_{k'} + V'_k V^*_{k'})\nonumber\right.\\ 
    &\,\,\left.-\left[\bk\cdot\bq\,f'_j\pol^\sigma_{j\bq} + \bk\cdot(\bq\bcurl\ez)\,f_j \tor^\sigma_{j\bq}\right](\hat{\bk}\cdot\hat{\bk}'\,H_k H^*_{k'} + V_k V^*_{k'})\right\}\Bigg]
\end{align}
We desire to linearly relate the coupling integral in the above equation to the flows $\pol$ and $\tor$, through poloidal and toroidal sensitivity kernels, $C_{\bq j,\bk}$ and $D_{\bq j,\bk}$ respectively. Hence, they are given by
\begin{eqnarray}\label{kernels}
&&C_{\bq j,\bk} = \int\textrm{d}z\, \rho \left[q^2\,f_j\,(\hat{\bk}\cdot\hat{\bk}'\,{H}'_k H^*_{k'} + V'_k V^*_{k'})\right.\nonumber\\ 
&&\left.-\bk\cdot\bq\, f'_j\, (\hat{\bk}\cdot\hat{\bk}'\,H_k H^*_{k'} + V_k V^*_{k'})\right],\nonumber\\
&&D_{\bq j,\bk} = \bk\cdot(\bq\bcurl\ez)\,\int \textrm{d}z\, \rho \, f_j\, (\hat{\bk}\cdot\hat{\bk}'\,H_k H^*_{k'} + V_k V^*_{k'}).
\end{eqnarray}
Note the symmetry $C_{\bq j,\bk} = C_{-\bq j,-\bk}$ and $\mathcal{D}_{\bq j,\bk}=\mathcal{D}_{-\bq j,-\bk}$. 
This coupling integral contributes to the cross-spectral measurement between modes $\bk$ and $\bk+\bq$ 
From eq 8 of \citet{Woodard14}, we write the first-order effect of flow on wavefield cross-correlation as 
\begin{equation}\label{cross correlation forward integral}
    \langle \phi^{\omega*}_\bk\,\phi^{\omega+\sigma}_{\bk+\bq} \rangle =
     H^\omega_{kk'\sigma}\Lambda^{\bk}_{\bk'}(\sigma),
\end{equation}
where the function $\mathcal{H}$ is given by 
\begin{eqnarray}\label{hfactor}
{\mathcal H}^\omega_{kk'\sigma} = -2\ii\omega(N_{k} |R^{\omega}_{k}|^2\, R^{\omega + \sigma}_{k'} + N_{k'} |R^{\omega+\sigma}_{k'}|^2\, R^{\omega*}_{k} ).
\end{eqnarray}
We absorb the factor $-2\ii\omega$ into the definition of $\mathcal{H}$. Substitute eq~\ref{kernels} in right-hand-side of eq~\ref{cross correlation forward integral} to obtain eq~\ref{cross-correlation forward model}. \\
The mode spectral profile $R$ is a Lorentzian, given by \begin{equation}\label{lorentzian}
    R^{\omega}_{k} = \frac{1}{\omega_{nk}^2-\omega^2-i\omega\gamma_{nk}/2},
\end{equation}
where $\omega_{nk}$ is the resonant frequency of the mode, and $\gamma_{nk}$ is the mode linewidth. 
Eq~\ref{lorentzian} can be derived by introducing mode damping $-\ii \omega \gamma \rho$ as an operator in the differential equation that governs undamped, driven oscillations \citep[see eq 5 of][]{Hanasoge17}, and then deriving the effects of first-order perturbations to the wavefield cross-correlation. Also, the parity ${\mathcal H}^\omega_{kk'\sigma} = {\mathcal H}^{-\omega*}_{kk'-\sigma}$ and $R^{\omega}_{k} = R^{-\omega*}_{k}$ are established.
Mode normalization  $N$ is given by 
\begin{equation}\label{normalization}
N_k = \frac{1}{\text{Q}}\sum\limits_{\bk}^{\text{Q}} \;\frac{\sum\limits_{\omega} \, |\phi_\bk^\omega|^2} {\sum\limits_\omega R^{\omega}_{k} },
\end{equation}
where the $\frac{1}{\text{Q}}\sum\limits_{\bk}^\text{Q}$ on the right-hand-side implies average over all $[k_x, k_y]$ (Q terms in all) such that $k=|\bk|$ is constant. This forces $N$ to be isotropic, i.e., to only depend on $k$, and not $\bk$. The sum over $\omega$ is within five linewidths of $\omega_{nk}$. Note that Eq.~\ref{hfactor} through~\ref{normalization} are modified from H21 to ensure parity and that flow maps are real.

The three equations~\ref{hfactor} through~\ref{normalization}, along with the symmetry relation for kernels, and summation over $\pm \omega$, serve to establish the parity $B_{\bk,\bq}^{\sigma} = B_{-\bk,-\bq}^{*-\sigma}$. This allows for obtaining $\pol_{\bq}^{\sigma} = \pol_{-\bq}^{*-\sigma}$, and subsequently, purely real flow in the real domain. Setting $\sigma=0$ gives us the linear, invertible equation eq~\ref{forward problem}. Substituting eqns~\ref{hfactor} through~\ref{normalization} into the noise model obtained in H21 and summing over $\pm \omega$ establishes the symmetry $G_{\bk,\bq}^{\sigma} = G_{-\bk,-\bq}^{-\sigma}$.

\section{SOLA inversions}\label{appendix b}
Subtractive Optimally Localized Averages \citep[SOLA,][]{Pijpers94} aims to obtain a set of weight factors for the mode $\bq$ and depth $z_o$, which we will call $\alpha_{\bk,zo}$. A linear weighted sum of the measurements $B_{\bk,\bq}$ in the fashion $\sum\limits_{\bk} \alpha_{\bk,zo} B_{\bk,\bq}$ allows for an average value of the flow $\pol_{\bq}(z)$ to be estimated at the depth $z_o$. To obtain the coefficients $\alpha_{\bk,zo}$, it is assumed that a set of sensitivity kernels $K_{\bk,\bq}(z)$ for the mode $\bq$ can be summed up coherently to give an 'averaging kernel' that is localized at the depth $z_o$. Conventionally, a Gaussian centered at $z_o$ and a width $\Delta$ is chosen which the averaging kernel should resemble after performing inversion.
\subsection{Kernels in the integral form}
\begin{figure}
\subfloat{\includegraphics[width = 6.2in]{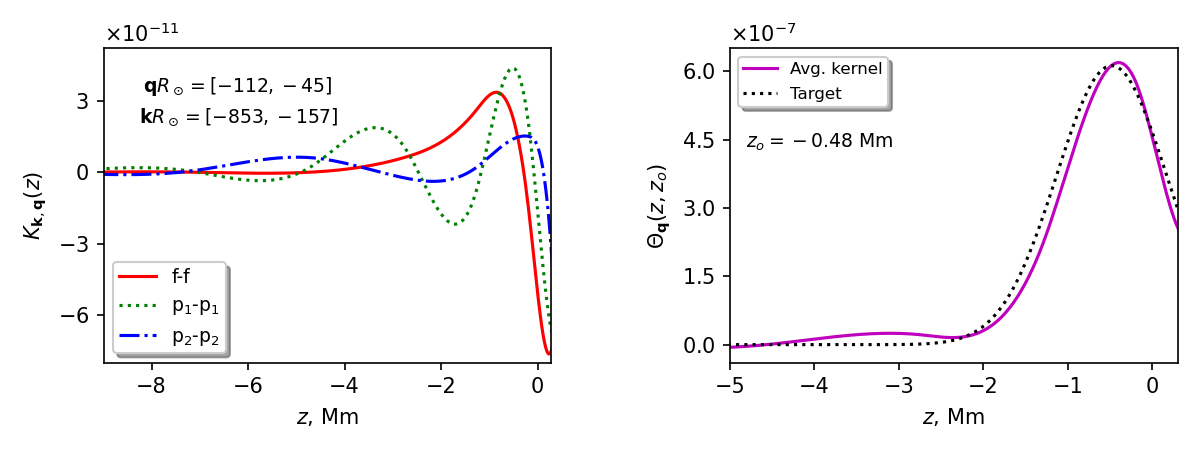}}
\caption{\textit{Left}: Kernel $K_{\bk,\bq}(z)$ (eq~\ref{kernel in integral}) shown vs depth $z$ for the three radial order couplings f-f, p$_1$-p$_1$, and p$_2$-p$_2$. $\bq R_\odot = [-112,-45]$ and $\bk R_\odot = [-853,-157]$ is chosen for all the radial order couplings for comparison. \textit{Right}: Averaging kernel (eq~\ref{avgkernel}) using SOLA, for $\bq R_\odot = [-112,-45]$ at depth $z_0 = -0.48$ Mm, and the corresponding target Gaussian (eq~\ref{target}). Integral of the averaging kernel over $z$ is 0.89.}
\label{fig_kernel}
\end{figure}

Since the kernels in eq~\ref{kernels} are manifest as coefficients on a basis $f_j(z)$, we first derive kernels that can be expressed as a function of depth $z$ (see Figure~\ref{fig_kernel}). It is convenient to derive in matrix form. Thus, with the following definitions: $P \equiv P_{\bq}(z)$, $p \equiv P_{\bq j}$, $F \equiv f_{j}(z)$, $B \equiv B_{\bk,\bq}$ $C \equiv C_{\bq j, \bk}$ and $K \equiv K_{\bk,\bq}(z)$, we write (assume only poloidal flow for simplicity, the same derivations hold true for toroidal flow as well)
\begin{equation}
    P = Fp
\end{equation}
The size of $P$ is thus the same as the length of the radial grid $z$.\\
Now, pre-multiply by $F^T$ and integrate over $z$ on both sides (drop the integral notation for compactness),
\begin{align}\label{poloidal vs z}
     F^{T}P &= (F^{T} F) p\nonumber\\
    p &=  (F^{T}F)^{-1} \,  F^{T}P
\end{align}
Now, substituting eq~\ref{poloidal vs z} into the forward problem eq~\ref{forward problem},
\begin{align}\label{forward problem integral}
    B &= Cp\nonumber\\
     &=(F^T F)^{-1} F^T C  P \nonumber\\
     & = KP
\end{align}
where
\begin{align}\label{kernel in integral}
    K &= (F^T F)^{-1}  F^T C, \nonumber\\
    \text{i.e.,} \;\;\;\; K_{\bk,\bq}(z) &= \sum\limits_{j,j'}\Big[\int \textrm{d}z\, f_j(z)f_{j'}(z)\Big]^{-1} f_{j'}(z) C_{\bq j',\bk}
\end{align}

\subsection{Obtaining the coefficients $\alpha$}\label{section alpha}
Now, demand that the averaging kernel should resemble a unimodulus target Gaussian centered at $z_o$
\begin{equation}\label{target}
\mathcal{T}(z,z_o) = \frac{1}{\sqrt{2\pi\Delta^2}} \;\text{exp}\Big(\frac{z-z_o}{2\Delta^2}\Big).    
\end{equation}
This can be achieved by solving the optimization problem 
\begin{align}
    {\text{minimize}} \; \mathcal{X} = \int \textrm{d}z \; \Big[\mathcal{T}(z,z_o) - \Theta_{\bq}(z,z_o) \Big]^2,
\end{align}
where we introduce the averaging kernel for mode $\bq$ thus 
\begin{equation}\label{avgkernel}
\Theta_{\bq}(z,z_o) = \sum\limits_{\bk} \alpha_{\bk,zo} K_{\bk,\bq}(z).    
\end{equation}
As an aside, we note that averaging kernels can similarly be constructed for RLS (see section~\ref{RLS}) using eqns~\ref{rlsinversion} and~\ref{kernel in integral}. 

\begin{figure}
\subfloat{\includegraphics[width = 5in]{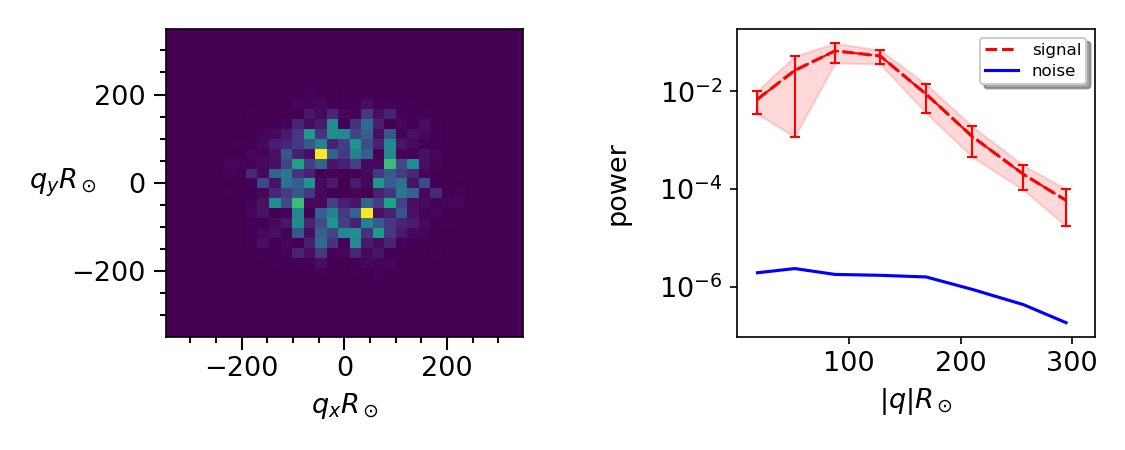}}
\caption{\textit{Left}: Poloidal flow power-spectrum for f-f as a function of $q_x R_\odot$ and $q_y R_\odot$. \textit{Right}: Corresponding power-spectrum averaged over the azimuthal angle. Shaded region shows $\pm1-\sigma$ error around the mean. Power is in units of m$^2$/s$^4$.}
\label{fig_sola_ps_azimuth}
\end{figure}

\begin{figure}
\subfloat{\includegraphics[width = 6.4in]{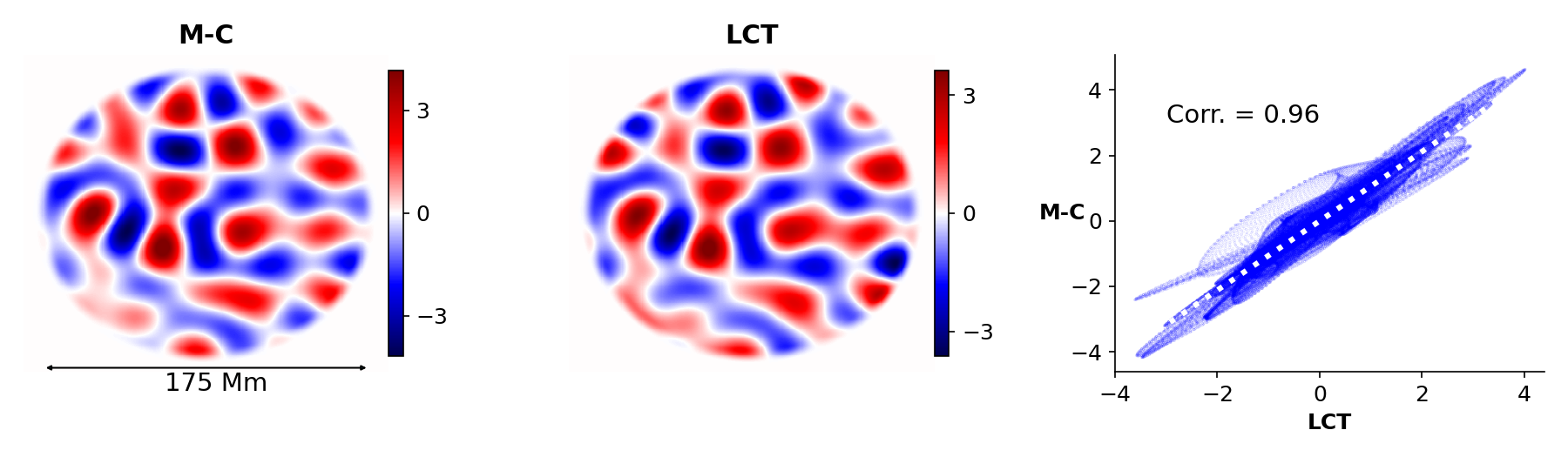}}
\caption{Real-space divergence flows (in units of $10^{-5}$s$^{-1}$) for mode-coupling inversion through SOLA using f-f coupling, and LCT, bandpass filtered around $qR_\odot=100$. We cut  edges out from the flow maps and compare a circular region of diameter $\approx$175 Mm. The scatter plot shows the agreement between the maps. The slopes of the best-fit line through the scatter plot is 1.05. For demonstration, we show inversions only for poloidal flow using SOLA.}
\label{fig_sola_real_space}
\end{figure}

Setting $\frac{\partial\mathcal{X}}{\partial\alpha}\rightarrow 0$ gives us the matrix problem to be solved 
\begin{align}\label{inverse sola}
    A\{\alpha\} &= v,\nonumber\\
    \{\alpha\} &= \Big[A + \mu I\Big]^{-1} v,
\end{align}
where the square matrix $A = \int\textrm{d}z\, K_{\bk,\bq}(z)K_{\bk',\bq}(z)$ and $v = \int\textrm{d}z\, K_{\bk,\bq}(z) \mathcal{T}(z,z_o)$. Here, $\bk'$ is just a dummy index for denoting elements in the matrix $A$, ($\bk'\neq\bk+\bq$). In the last line of eq~\ref{inverse sola}, we introduce regularization using an Identity matrix $I$, with the regularization parameter $\mu$ - purpose being the same as that described in section~\ref{RLS}.
Obtaining $\alpha$ thus becomes a highly expensive computationally for very large number of modes (see section~\ref{inversion}). Substitute $\alpha$ obtained from eq~\ref{inverse sola} into last line of eq~\ref{forward problem integral}, and $\sum\limits_{\bk}$ on both sides
\begin{align}
    \sum\limits_{\bk} \alpha_{\bk,z_o} B_{\bk,\bq}^{\sigma} &= \sum\limits_{\bk}\alpha_{\bk,z_o}\;\int\textrm{d}z\, K_{\bk,\bq}(z) P_{\bq}^{\sigma}(z),\nonumber \\
    &= \int\textrm{d}z\,\Theta_{\bq}(z,z_o) P_{\bq}^{\sigma}(z),\nonumber\\
    &\approx \langle P_{\bq}^{\sigma}(z_o)\rangle
\end{align}
Inversions can similarly be performed for multiple depths by choosing suitable widths for the target Gaussians. Divergence flow can then be obtained from eq~\ref{mcdiv}. Results are shown in Figures~\ref{fig_sola_ps_azimuth} and~\ref{fig_sola_real_space}.

\bibliography{References}{}
\bibliographystyle{aasjournal}
\end{document}